\documentclass[twocolumn,english,prb,aps,amssymb,superscriptaddress,showpacs]{revtex4-1}
\usepackage[T1]{fontenc}
\usepackage{amsmath}
\usepackage{amssymb}
\usepackage{graphicx}

\makeatletter
\@ifundefined{textcolor}{}
{%
 \definecolor{BLACK}{gray}{0}
 \definecolor{WHITE}{gray}{1}
 \definecolor{RED}{rgb}{1,0,0}
 \definecolor{GREEN}{rgb}{0,1,0}
 \definecolor{BLUE}{rgb}{0,0,1}
 \definecolor{CYAN}{cmyk}{1,0,0,0}
 \definecolor{MAGENTA}{cmyk}{0,1,0,0}
 \definecolor{YELLOW}{cmyk}{0,0,1,0}
}

\def\be{\begin{equation}}
\def\ee{\end{equation}}
\def\bea{\begin{eqnarray}}          
\def\eea{\end{eqnarray}}
\def\bi{\begin{itemize}}
\def\ei{\end{itemize}}

\usepackage{epsfig}
\usepackage{dcolumn}
\usepackage{bm}

\makeatother

\usepackage{babel}
\begin{document}

\title{ 
           Striped critical spin liquid in a spin-orbital entangled RVB state \\
                in a projected entangled-pair state representation 
}

\author{Piotr Czarnik}
\author{Jacek Dziarmaga}
\affiliation{Instytut Fizyki Uniwersytetu Jagiello\'nskiego,
             ul. Reymonta 4, PL-30059 Krak\'ow, Poland}

\date{June 20, 2014}

\begin{abstract}
We introduce a spin-orbital entangled (SOE) resonating valence bond (RVB) state on a square lattice of spins-$\frac12$ and orbitals represented by pseudospins-$\frac12$. 
Like the standard RVB state,
it is a superposition of nearest-neighbor hard-core coverings of the lattice by spin singlets, 
but adjacent singlets are favoured to have perpendicular orientations and, 
more importantly, 
an orientation of each singlet is entangled with orbitals' state on its two lattice sites. 
The SOE-RVB state can be represented by a projected entangled pair state (PEPS) with a bond dimension $D=4$. 
This representation helps to reveal that the state is a superposition of striped coverings conserving a topological quantum number. 
The stripes are a critical quantum spin liquid. 
We propose a spin-orbital Hamiltonian supporting a SOE-RVB ground state. 
\end{abstract}


\maketitle

\section{Introduction}

Spin-orbital interplay is one of the most important topics in the theory of frustrated magnetism 
\cite{Nag00,Hfm,Kha05,Ole05,Ole12}. When degenerate $3d$ orbitals in a transition metal oxide are 
partly filled, realistic superexchange includes both orbital and spin degrees of freedom that are 
strongly interrelated \cite{Kug82,Fei97}. In many cases, the intertwined spin-orbital interaction 
may be decoupled by mean field approximation, and the spin and orbital dynamics are independent 
from each other. Thus a spin-only Heisenberg model can be derived by averaging over the orbital 
state, which successfully explains magnetism and optical excitations in some materials, as for 
instance in LaMnO$_3$ \cite{Fei99}. Recent interest and progress in the theory of spin-orbital 
superexchange models was triggered by the observation that orbital degeneracy drastically increases 
quantum fluctuations which may suppress long-range order in the regime of strong competition 
between different types of ordered states near a quantum critical point \cite{Fei97,Brz12}. 
Spin and orbital variables are here entangled.

The topological spin liquids (TSL) are at the forefront of condensed matter theory and quantum 
information \cite{WhiteKagome}. They serve as an example of strongly correlated systems with 
non-Landau non-local order parameters. Their non-Abelian excitations can be used to operate 
a topological quantum computer \cite{topoc,Tro10}. There are exactly solvable models with TSL 
ground states \cite{Kit06}. The search for realistic models gained momentum after White demonstrated 
the TSL nature of the Kagome antiferromagnet \cite{WhiteKagome}. This result was obtained by 
a tour-de-force application of the quasi-1D density matrix renormalization group, a technique 
elevated to a higher degree of sophistication in Ref. \cite{CincioVidal}. However, the use of 
DMRG in 2D is limited to systems with short-range correlations only, a restriction that does 
not apply to the PEPS ansatz \cite{PEPS} whose usefulness for TSL has already been demonstrated 
\cite{PepsRVB,PepsKagome,Wan13}, including doped RVB states \cite{dopedRVB}. 

In this paper we introduce a spin-orbital entangled resonating valence bond (SOE-RVB) state on 
a square lattice with a spin-$1/2$ and orbital degrees of freedom represented by a pseudospin-$1/2$ 
at every lattice site. An orientation of each nearest-neighbor spin singlet along one of the lattice 
axes is entangled with the orbitals on its two lattice sites. The adjacent singlets are preferred 
to have perpendicular orientations as is often the case in the spin-orbital systems \cite{Brz12}. 
We use a PEPS representation of the SOE-RVB state to demonstrate that it is a striped topological 
quantum spin liquid with critical correlations.

The paper is organized as follows. 
In Section \ref{sec:soervb}, we introduce the SOE-RVB state on a square lattice. 
In Section \ref{sec:PEPS}, 
we construct the PEPS representation of this state. 
In Section \ref{sec:transfer},
the lattice is compactified to a cylinder of a finite circumference $L$ 
and a transfer matrix along the cylinder is defined. 
In Section \ref{sec:diagonal}, 
we introduce a simplification of the transfer matrix that is justified in the thermodynamic limit $L\to\infty$.
In Section \ref{sec:stripes},
the SOE-RVB state is shown to be a superposition of striped coverings that preserve a topological quantum number. 
In Section \ref{sec:correlations},
correlations in the striped coverings are found to decay algebraically with a distance like in a critical state.
In Section \ref{sec:simpleH}, 
we introduce a toy spin-orbital Hamiltonian supporting the SOE-RVB state as a ground state, 
and a weak perturbation of the toy model that removes the degeneracy between different topological quantum numbers.
In Section \ref{sec:KKH},
a realistic Kugel-Khomskii Hamiltonian is shown to have the same energy in the SOE-RVB state as in the plaquette RVB state proposed earlier. 
Finally, we conclude in Section \ref{sec:conclusion}. 
Some technical details were left for the appendix.

\section{SOE-RVB state}\label{sec:soervb}

At each lattice site of a 2D square lattice, in addition to a spin-$\frac12$, there is a pseudospin-$\frac12$ representing 
orbital degrees of freedom. The spin-orbital RVB state is a weighted sum over spin-singlet coverings
\be
\sum_{C} w_C(\theta) |C\rangle.
\ee
It runs over all nearest-neighbor (NN) hard-core coverings, where each lattice site is covered by a NN spin-singlet.
Each singlet is oriented from a sublattice $A$ to $B$: $\left(|0_A\rangle|1_B\rangle-|1_A\rangle|0_B\rangle\right)/\sqrt2$. 
However, this state differs from the usual spin RVB state in two important respects:
\bi 
\item a horizontal(vertical) spin singlet is associated with an orbital state $|0_A0_B\rangle$($|1_A1_B\rangle$) on its two sites;
\item the weights prefer pairs of adjacent singlets to be perpendicular to each other.
\ei
The entanglement between the singlet orientation and the orbitals' polarization is enough to make any two coverings that differ 
by just one pair of overlapping perpendicular singlets mutually orthogonal. The bias toward perpendicular adjacent singlets
is effected by the weights $w_C(\theta)=\cos^K\theta$, where the parameter $\theta\in[0,\pi/2)$ and $K$ is a number of NN lattice 
bonds connecting parallel singlets. When $\theta=0$ we recover the standard orthogonal dimer state, but here we are more interested 
in the opposite limit $\theta\to\pi/2$, when parallel singlets are suppressed as much as possible, even though they cannot be 
quite eliminated. We study this limit in the PEPS representation. 
 
\section{PEPS representation}\label{sec:PEPS}

\begin{figure}[t]
\begin{center}\includegraphics[width=8cm]{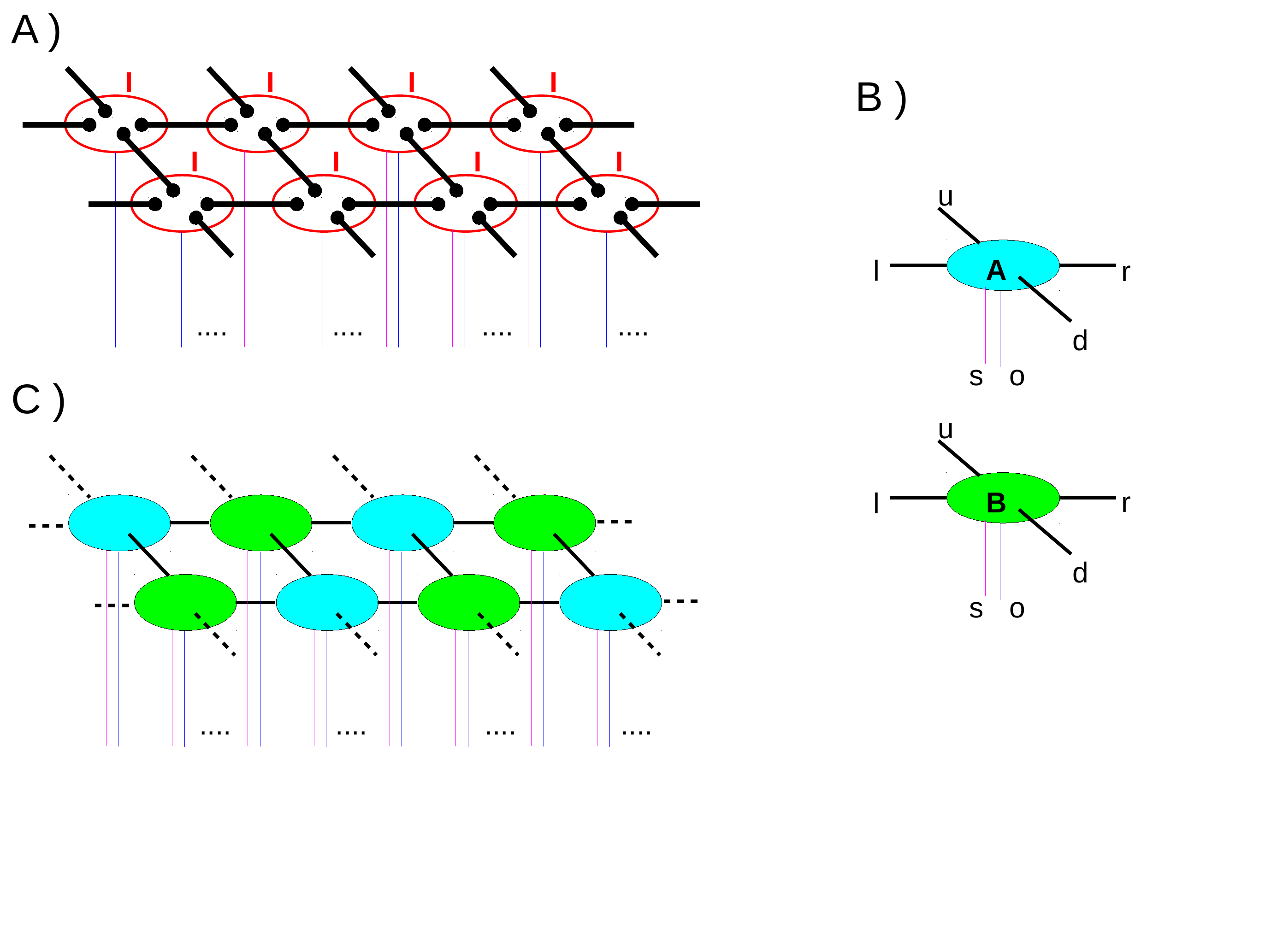}\end{center} 
\vspace{-1.7cm}
\caption{
In A, 
the entangled-pair state (EPS) in (\ref{entangl}) is placed on each bond connecting NN lattice sites.
Then the isometry (projector) $Z$ in (\ref{Z}) is applied at each site.
In B,
the projection creates at each site of the sublattice $A$ or $B$ a tensor $A$ or $B$ respectively.
In C,
the tensors contracted through their bond indices, 
but with free spin and orbital indices, 
make the PEPS representation of the SORVB state. 
}
\label{FigPeps} 
\end{figure}

We define two states in an auxiliary Hilbert space:
\begin{eqnarray}
|a) &=& |2)\cos\frac{\theta}{2}+|3)\sin\frac{\theta}{2}, \nonumber\\
|b) &=& |2)\cos\frac{\theta}{2}-|3)\sin\frac{\theta}{2},
\label{ab}
\end{eqnarray}
with a product $(a|b)=\cos\theta$. On each bond connecting NN sites we place a virtual entangled-pair state (EPS)
\begin{equation} 
|0_A1_B)-|1_A0_B)+|2_A2_B)+|3_A3_B),
\label{entangl}
\end{equation}
see Figure \ref{FigPeps}A. Then at each lattice site we apply an isometry (projector)
\begin{eqnarray}
I=
&\sum_{i=0,1}&
|i_S0_O\rangle
\left[
(a_U i_R b_D a_L|+
(a_U b_R b_D i_L|
\right]+  \nonumber\\
&&
|i_S1_O\rangle
\left[
(b_U a_R i_D b_L|+
(i_U a_R a_D b_L|
\right]. 
\label{Z}
\end{eqnarray}
Here $S,O$ refer to the spin and orbital on the site and $U,R,B,L$ to the bonds coming out from it:
$L,R$ along the $a$-axis and $U,D$ the $b$-axis. A resulting projected-EPS (PEPS) tensor 
on the sublattice $A$ is
\begin{eqnarray} 
A^{s,o}_{u,r,d,l}
&=&
\delta_{o0}
(u|a)(d|b)
\left[
(l|a) \delta_{rs} + \delta_{ls} (r|b)
\right]+
\nonumber\\
&&
\delta_{o1}
(l|b)(r|a)
\left[
\delta_{us} (d|a) + (u|b) \delta_{ds}
\right]
\label{ARVB}
\end{eqnarray}
Here the bond indices $u,r,d,l\in\{0,1,2,3\}$ (the PEPS bond dimension $D=4$) 
and the spin and orbital indices $s,o\in\{0,1\}$. 
A PEPS tensor on the sublattice $B$ is 
\begin{equation}
B^{s,o}_{u,r,d,l}=
(-1)^s \sum_{s'} \sigma^x_{s,s'} A^{s',o}_{u,r,d,l}.
\label{BRVB}
\end{equation}
A director $|a\rangle$ or $|b\rangle$ sticking out of a PEPS tensor along one of its bonds signals to 
its environment the orientation of a singlet covering its site. Each bond connecting parallel singlets 
is suppressed by a factor $(a|b)=\cos\theta$.

\section{Transfer matrix}\label{sec:transfer}

\begin{figure}[t]
\begin{center}\includegraphics[width=8cm]{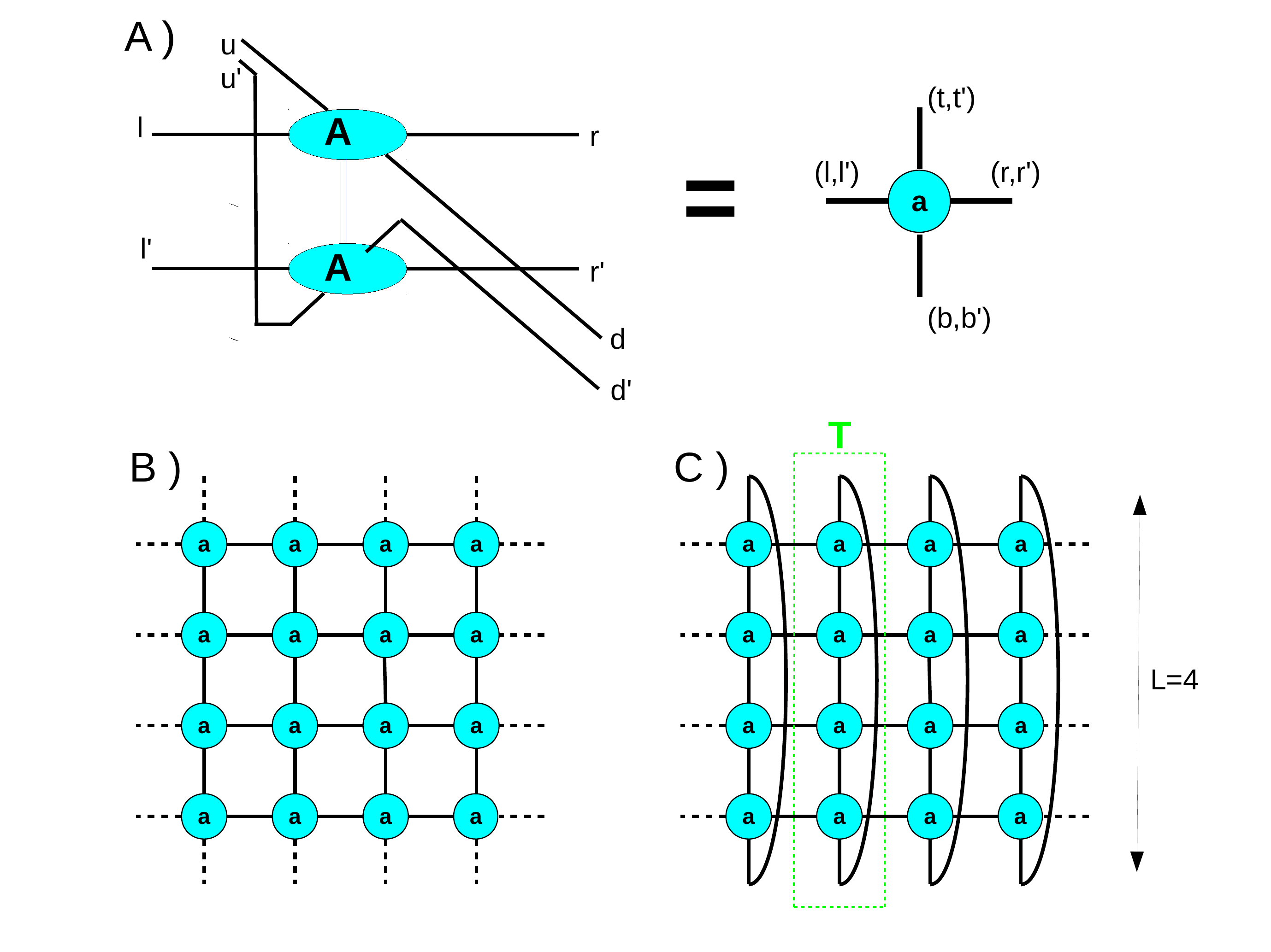}\end{center} 
\vspace{-1.0cm}
\caption{
In A,
a contraction of the PEPS tensor $A$ with its conjugate through the spin and orbital indices makes a transfer tensor $a$.
The same $a$ is obtained on the sublattice B after contraction of the tensor $B$ with its conjugate. 
In B,
a norm squared of the PEPS is represented by a contraction of the transfer tensors. 
In C,
the norm squared of the PEPS on a periodic cylinder of circumference $L=4$.
A contraction of all transfer tensors $a$ in a periodic column makes a transfer matrix $T$.
}
\label{FigAA} 
\end{figure}

To calculate its expectation value an operator has to be sandwiched between two PEPS's in Figure (\ref{FigPeps})C.
In the ``sandwich'' each PEPS tensor on a site that does not belong to the support of the operator 
is contracted with its conjugate through their spin and orbital indices, see Figure \ref{FigAA}A. 
This contraction makes a transfer tensor $a$. 
Comparing (\ref{ARVB}) and (\ref{BRVB}) one finds that the transfer tensor is the same on both sublattices. 
The norm squared of the PEPS is represented in Figure \ref{FigAA}B.
In Figure \ref{FigAA}C, the same norm is shown on a periodic cylinder of circumference $L$. 
A contraction of $L$ transfer tensors in each periodic column makes a transfer matrix $T$.

\section{Diagonal ensemble}\label{sec:diagonal}

The norm squared in Fig. \ref{FigAA}C is a sum 
$
\sum_{C,C'} w_C(\theta)w_{C'}(\theta) \langle C | C' \rangle.
$
For the overlap $\langle C | C' \rangle$  to be $0$ it is enough that the coverings 
$C$ and $C'$ differ by just a single pair of perpendicular SOE singlets that overlap on 
a common site.   
The only way that two different coverings can have a non-zero overlap is that
both coverings have a 1D train of singlets along the same line of sites, 
but the trains are shifted with respect to each other by one lattice site. 
However, with increasing system size the overlap between the shifted trains tends 
to zero exponentially fast. Since we are primarily interested in an extrapolation to 
the thermodynamic limit, in the following we approximate
\be 
\sum_{C,C'} w_C(\theta)w_{C'}(\theta) \langle C | C' \rangle
\approx
\sum_C w^2_C(\theta)
\ee
ignoring any overlaps between different coverings.

This ``diagonal'' approximation simplifies radically the transfer tensor $a$ in Figure \ref{FigAA}A. 
Instead of $D^2=16$ different values, each of its four bond indices can take only $3$ values numbering 
an orthonormal basis in a subspace spanned by the following states 
\be 
|aa),~|bb),~\left(|00)+|11)\right)/\sqrt2
\ee
with the last one indicating a singlet along a given bond. 
The three states become orthonormal when $\theta\to\pi/2$.
The reduced bond dimension accelerates contraction of tensor networks like the one in Fig. \ref{FigAA}C.

Some further reductions follow from the fact that the singlet can stick out only from one of the four indices 
of the transfer tensor $a$, i.e., if one index shows the singlet, then the other indices must show either 
$|aa)$ or $|bb)$. These correlations help to compactify the column transfer matrix $T$.

\section{Topological stripe coverings}\label{sec:stripes}

\begin{figure}[t]
\begin{center}\includegraphics[width=8cm]{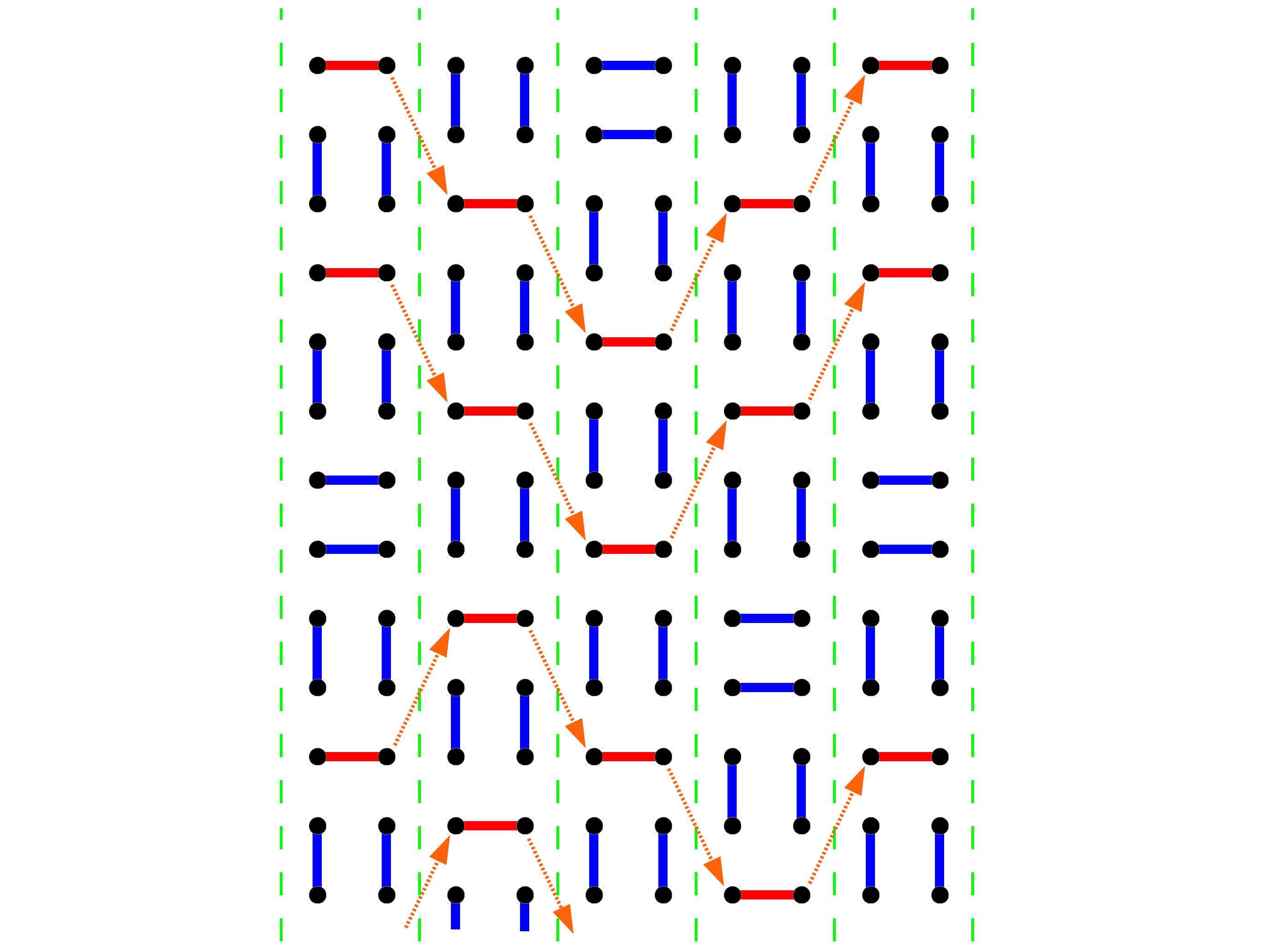}\end{center} 
\vspace{-0.5cm}
\caption{
A typical singlet covering in the striped spin liquid. 
Here the black dots are lattice sites and the thick connecting lines represent spin singlets. 
The green dashed lines separate different stripes.
The unpaired singlets are highlighted red.
The orange arrows indicate their jumps between NN stripes. 
}
\label{stripes} 
\end{figure}

The tensor network in Fig. \ref{FigAA}C, corresponding to an infinite cylinder, 
is a graphic representation of an infinite power of the transfer matrix: $T^N$ with $N\to\infty$.
When a finite segment of length $l$ of the infinite cylinder is concerned, 
then instead of the infinite power one can consider a finite object $(L|T^l|R)$,
where $\left(L\right|$ and $\left|R\right)$ are respectively the left and right dominant eigenvectors of $T$.
The compact object $(L|T^l|R)$ can be used to scrutinize directly what singlet coverings appear in the finite
segment and what are their probabilities.   
The direct scrutiny was completed up to $L=12$.
In the limit $\theta\to\pi/2$ only striped coverings survive, 
like the typical example in Fig. \ref{stripes}, 
with either vertical or horizontal orientation of the stripes.
The following Table shows a relative weight of vertical stripes versus all coverings as a function of 
$\theta$ for $L=11$. 

\vspace{-0.2cm}
\begin{center}
\begin{tabular}{ |c|c| }
\hline
$\theta/\frac{\pi}{2}$ & weight \\
\hline
$0.70$ & 0.5002\\
$0.80$ & 0.7959\\
$0.90$ & 0.9809\\
$0.92$ & 0.9920\\
$0.95$ & 0.9988\\
\hline
\end{tabular} 
\end{center}  
\vspace{-0.2cm}      
   
For the sake of definiteness, 
in the following we focus on vertical stripes like in Fig. \ref{stripes}.
The minimization of the number of bonds connecting parallel singlets in the limit $\theta\to\pi/2$ leads to an interesting topological conservation law. 
All singlets appear in parallel pairs, 
except for some unpaired horizontal singlets. 
The number of unpaired singlets in each vertical stripe is the same. 
This is a topologically protected quantum number. 
Only locations of singlets can change when passing between NN vertical stripes, 
but their total number remains the same. 
Each singlet jumps two lattice sites either up or down along a stripe, 
but with a constraint that two unpaired singlets cannot end at NN locations
or, equivalently, they cannot be swapped along the stripe. 
Locations of consecutive unpaired singlets along a stripe differ by $3+4j$ sites with $j=0,1,2,...$. 

\section{Correlations}\label{sec:correlations}

Since a vertical (horizontal) singlet is associated with the orbital state $|11\rangle$ ($|00\rangle$),
then an orbital operator 
\be 
Z=|1\rangle\langle1|-|0\rangle\langle0|
\ee
can be conveniently used to characterize singlet correlations. Its average, 
\be 
\langle Z \rangle = {\cal N}/L,
\ee 
is a fraction of unpaired singlets. The following Table shows its convergence with 
$\theta\to\frac{\pi}{2}$ for ${\cal N}=2,L=6$.

\vspace{-0.10cm}
\begin{center}
\begin{tabular}{ |c|c| }
\hline
$\theta/\frac{\pi}{2}$ & $\langle Z \rangle$ \\
\hline
$0.90$ & 0.333040\\
$0.95$ & 0.333315\\
$0.97$ & 0.333331\\
\hline
\end{tabular} 
\end{center}  
\vspace{-0.10cm}      

Correlators $\langle Z_i Z_j \rangle$ can be obtained either from the full transfer matrix $T$ or the striped coverings only. 
With the last method one can reach longer $L$. 
When the unpaired singlets are represented by hard-core bosons, 
then the ``striped'' transfer matrix for ${\cal N}$ particles becomes
\be 
T_{\rm S}~=~P~:\left( \sum_{l=1}^L c^\dag_{l+2} c_l + {\rm h.c.} \right)^{\cal N}:~P.
\label{TS}
\ee 
Here $l$ is a location along a periodic stripe, $c_l$ is a hard-core bosonic annihilation operator,
and $P$ is a projector enforcing the constraint that consecutive occupied locations differ by $3+4j$. 
An application of $T_{\rm S}$ to a Fock state with ${\cal N}$ particles on $L$ sites creates 
a superposition of Fock states with each particle shifted by $\pm2$ sites with respect to its original location.

\begin{figure}[t]
\begin{center}\includegraphics[width=8cm]{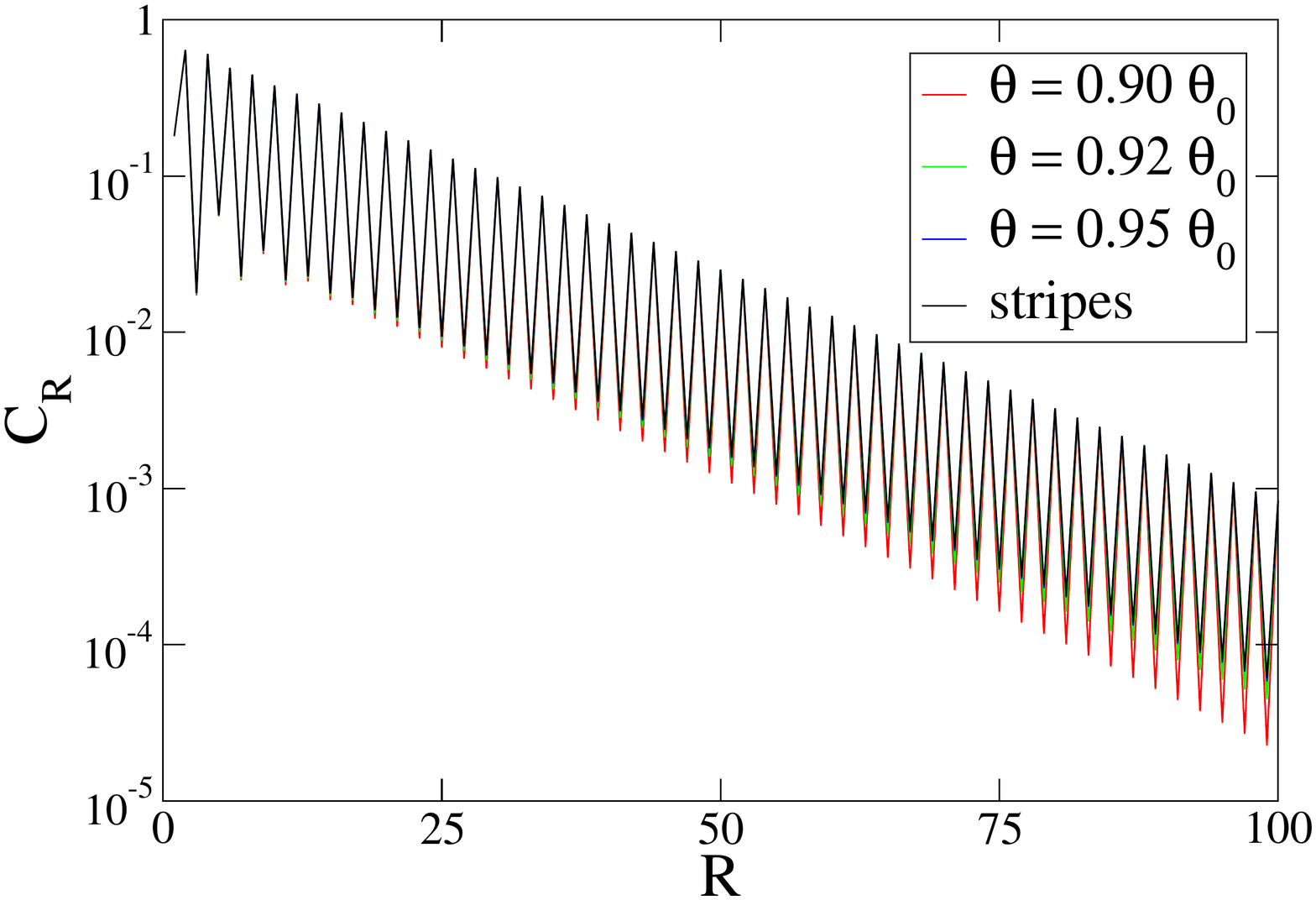}\end{center} 
\vspace{-0.8cm}
\caption{
The absolute value of the correlator $C_R$ in (\ref{CR}) along a cylinder with a circumference $L=10$.
With $\theta\to\frac{\pi}{2}$ the correlator obtained with the full transfer matrix $T$ converges 
to the correlator in the striped state obtained with $T_{\rm S}$. Here the dominant eigenvector 
of $T_{\rm S}$ has ${\cal N}=2$ unpaired singlets. The exponential decay of $C_R$ with $R$ can 
be fitted with a correlation length $\xi\approx15$. 
}
\label{CRL10} 
\end{figure}

\begin{figure}[t]
\begin{center}\includegraphics[width=8cm]{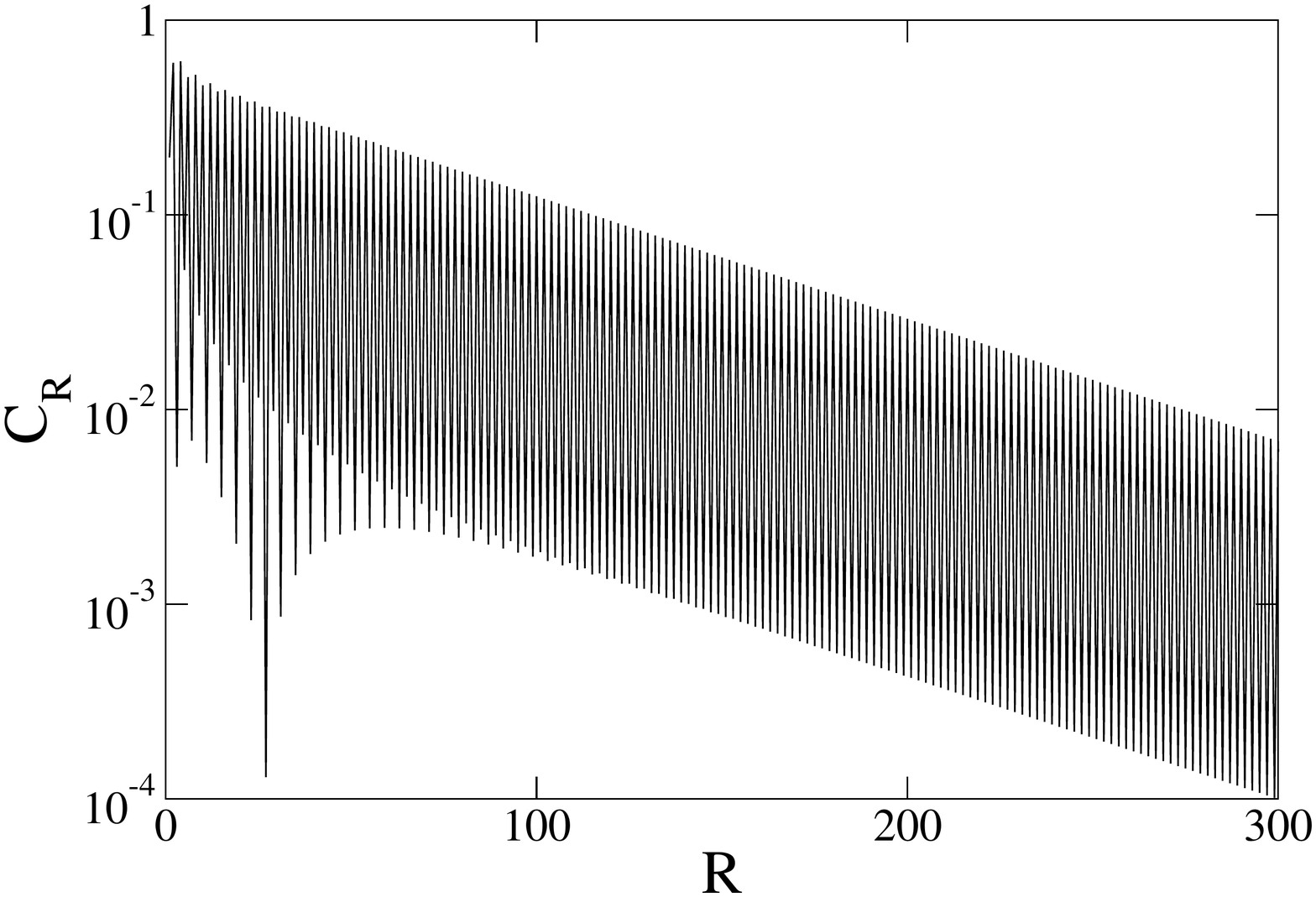}\end{center} 
\vspace{-0.8cm}
\caption{
The absolute value of the correlator $C_R$ in (\ref{CR}) along a cylinder with circumference $L=72$
obtained with $T_{\rm S}$ in the striped state. Here the dominant eigenvector of $T_{\rm S}$ 
has ${\cal N}=16$ unpaired singlets. The exponential decay of $C_R$ with $R$ can be fitted 
with a correlation length $\xi\approx70$. 
}
\label{xiL72} 
\end{figure}

\begin{figure}[t]
\begin{center}\includegraphics[width=8cm]{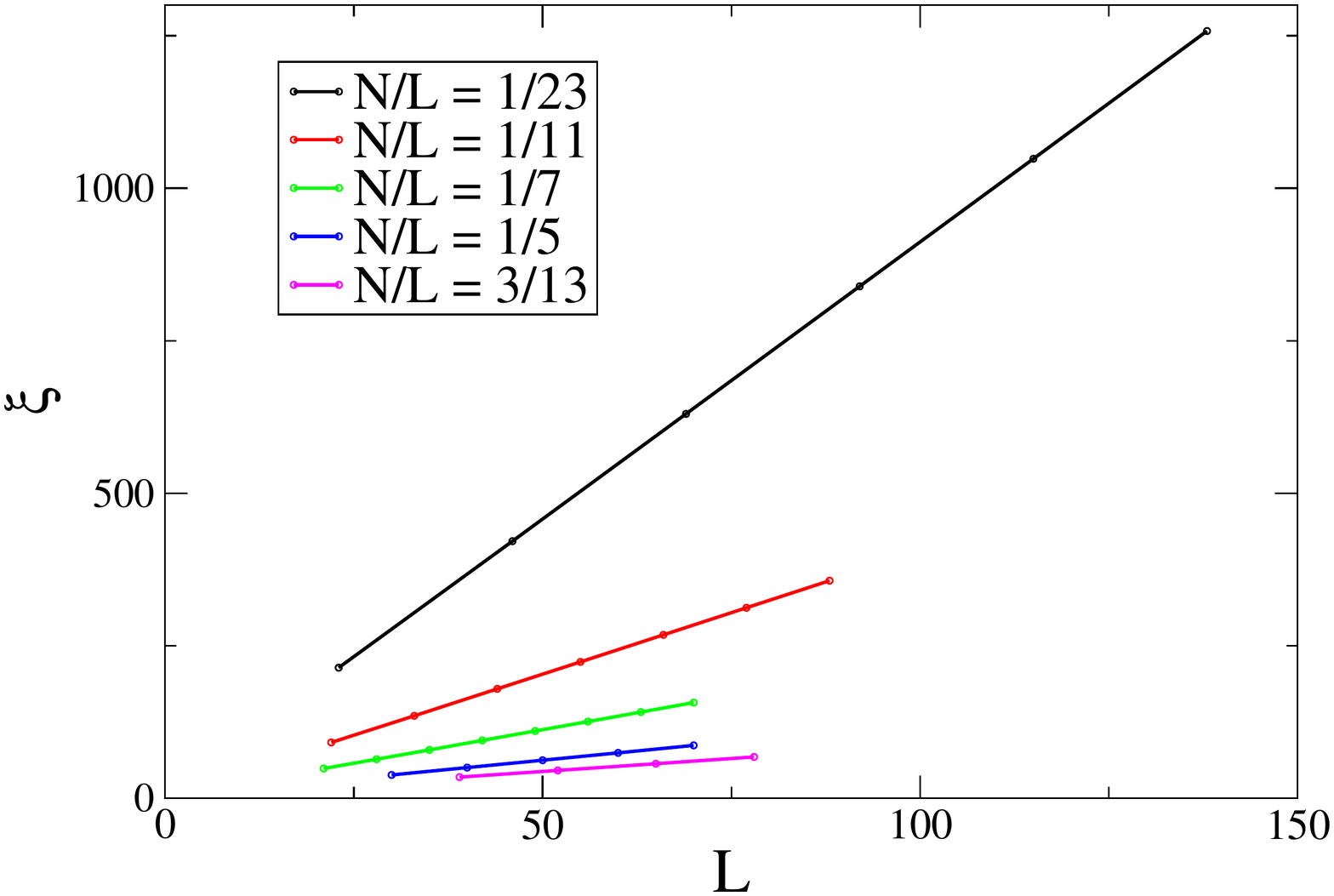}\end{center} 
\vspace{-0.8cm}
\caption{
The correlation length $\xi$ of the correlator $C_R$ along a cylinder as a function 
of the cylinder's circumference $L$ obtained with $T_{\rm S}$ in the striped state
for different values of the density ${\cal N}/L$ of unpaired singlets. 
$\xi(L)$ proves to be a linear function. 
Its slope decreases with the density. 
The linearity proves that the transfer matrix $T_{\rm S}$ has a gap between its two most dominant eigenvalues that decays like $1/L$.
The gap increases with the density of unpaired singlets.
}
\label{xiL} 
\end{figure}

Figure \ref{CRL10} shows a correlator
\be 
C_R = \left\langle Z_1 Z_{1+R} \right\rangle - \left\langle Z \right\rangle^2
\label{CR}
\ee
in the direction perpendicular to the stripes along a cylinder with a circumference $L=10$. When 
$\theta\to\frac{\pi}{2}$ the correlators obtained with the full transfer matrix $T$ converge to the 
correlator in the striped phase obtained with $T_{\rm S}$. Figure \ref{xiL72} shows the same correlator 
in the striped state along a cylinder with a large circumference $L=72$. 
For large $R$ both correlators in Figs. \ref{CRL10} and \ref{xiL72} decay exponentially, 
but with a correlation length $\xi$ that depends on $L$. 
In Figure \ref{xiL} we make a systematic study of this size dependence. 
$\xi$ proves to be a linear function of $L$ whose slope decreases with increasing density of unpaired singlets ${\cal N}/L$. 
We can conclude that in the thermodynamic limit $L\to\infty$ the decay of correlations in the direction perpendicular to the stripes is slower than exponential. 
Furthermore, since $\xi^{-1}$ is a gap between the two most dominant eigenvalues of $T_{\rm S}$, 
then this gap decays like $1/L$. 
The ``Hamiltonian'' $T_{\rm S}$ has ``low energy'' excitations with a linear dispersion relation.

The real symmetric $T_{\rm S}$ can be normalized to have the dominant eigenvalue $\Lambda_0=1$. 
With a spectral representation $T_{\rm S}=\sum_{m} \Lambda_m |m)(m|$ the correlator in the direction perpendicular to the stripes becomes
\be
C_{2R+2}^{\perp} = (0|Z T_{\rm S}^{R}  Z|0) - (0|Z|0)^2 = \sum_{m>0} (0|Z|m)^2 \Lambda_m^R.
\ee
For the gapless excitations we have $|\Lambda_m|\approx1-\lambda_m/L$ and, for a large $R$, 
$\Lambda_m^R\approx (\pm1)^Re^{-\lambda_mR/L}$.
Taking $L\to\infty$ and replacing the sum over $m$ with an integral over $\lambda$ we obtain
\be
C_{2R+2}^{\perp} \approx \int_0^\infty \frac{d\lambda}{L}~\sigma(\lambda/L)~z(\lambda/L)~e^{-\lambda R/L}.
\ee
Here $\sigma(..)$ is density of states and $z(..)$ is an average value of $(0|Z|m)^2$.
For a large $R$ when small $\lambda/L$ dominate
\be
C_{2R+2}^{\perp} \sim R^{-(\gamma+1)}.
\ee
Here $\gamma$ is an exponent in $\sigma(\lambda/L)z(\lambda/L)\sim (\lambda/L)^\gamma$ for small $\lambda/L$.

\begin{figure}[t]
\begin{center}\includegraphics[width=8cm]{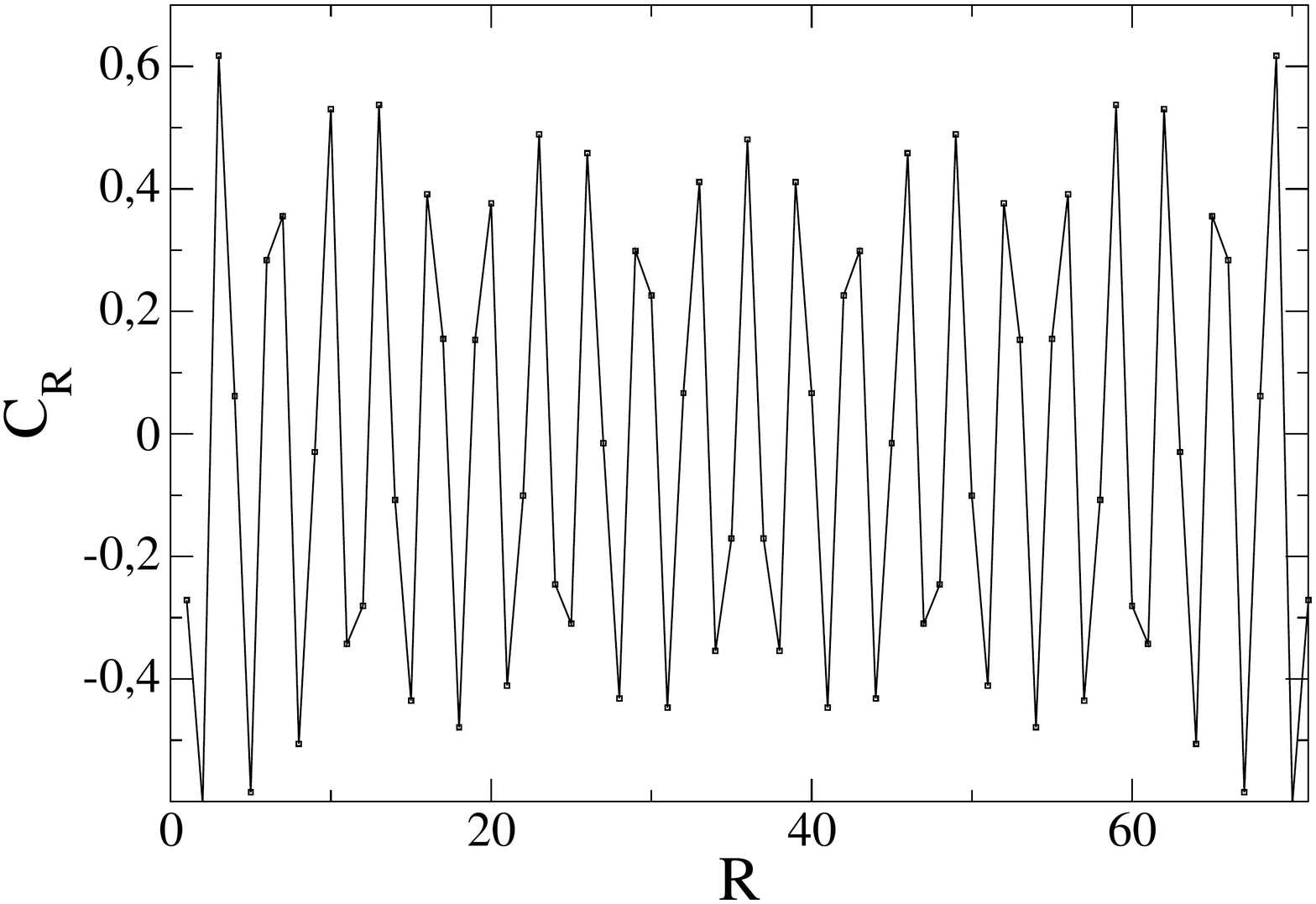}\end{center} 
\vspace{-0.8cm}
\caption{
The correlator $C_R$ in (\ref{CR}) along the stripes of length $L=72$ with the dominant ${\cal N}=16$.
The correlator's decay with $R$ is not significant.
}
\label{XXvert} 
\end{figure}

A numerical study in Figure \ref{XXvert} shows that the same correlator but in the direction along a stripe does not decay 
much within the numerically achievable distance. This is not quite unexpected given the gapless spectrum of $T_{\rm S}$ that implies 
a power-law decay of the correlator. This critical decay can be better understood in the regime of low density $\rho$ of unpaired 
singlets, where the detailed microscopic form of the constraint $P$ is irrelevant and the singlets can be replaced by 
a Tonks gas of impenetrable hard-core bosons. Furthermore, the correlator $C_R$ between sites, say, $l$ and $l+R$ depends 
only on a number of hard-core bosons on the sites $l+1,...,l+R-1$. A probability distribution for this number can be equivalently, 
and more conveniently, obtained from the Fermi sea with the same density of particles. As shown in detail in the Appendix,
its variance is logarithmic in $R$, hence $C_R$ decays as a power law: 
\be 
C_R\sim (\rho R)^{-1/8}\cos\frac{\pi}{2}(\rho+1)R.
\ee
The small exponent may explain the modest decay observed in Fig. \ref{XXvert}.

\section{Simple Hamiltonian}\label{sec:simpleH}

Let us consider a toy spin-orbital Hamiltonian
\begin{equation}
H_0=
\lambda\sum_{\langle i,j\rangle}Z_iZ_j+
\sum_{\langle i,j\rangle\parallel\gamma}
{\bf S}_i {\bf S}_j \left(\frac{1\mp Z_i}{2}\right) \left(\frac{1\mp Z_j}{2}\right).
\label{Htoy}
\end{equation}
Here ${\bf S}_i=\frac12{\bf\sigma}_i$ are spin operators and $Z_i$ orbital Pauli matrices. 
The upper(lower) signs correspond to a bond along $\gamma=a$($b$). Note that $Z_i$ are 
good quantum numbers.

When $\lambda<\frac34$ the energy of an isolated bond is minimized by a product of a spin singlet 
and an orbital $ZZ$-ferromagnet: $|00\rangle$ for a bond along $a$ and $|11\rangle$ along $b$.
This is just the SOE singlet in the SOE-RVB state.

\begin{figure}[t]
\begin{center}\includegraphics[width=8cm]{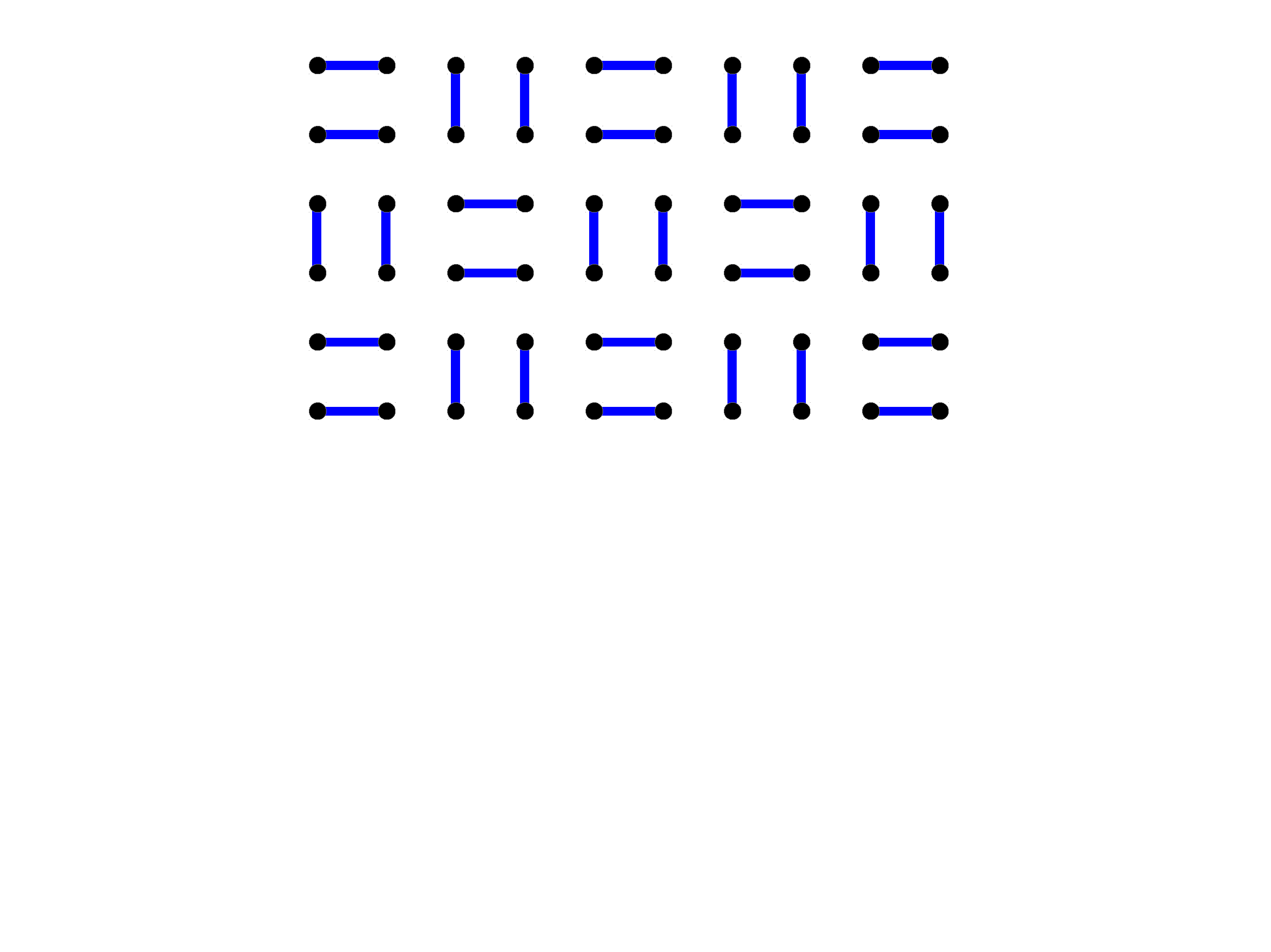}\end{center} 
\vspace{-3.3cm}
\caption{
The plaquette RVB (PVB) state \cite{Brz12}.
This is the special stripe covering without any unpaired singlets, ${\cal N}=0$.
}
\label{pvb} 
\end{figure}

Back on the lattice, we need $\lambda<\frac{3}{16}$ for the regular PVB covering in Figure \ref{pvb}
to have lower energy than the orbital Neel state. In the PVB state the adjacent singlets were made 
perpendicular as often as possible to minimize the energy of the $\lambda$-coupling between them. 
What is more, PVB is degenerate with all stripe coverings like the one in Fig. \ref{stripes}. It is 
the special stripe covering without any unpaired singlets, ${\cal N}=0$.

However, for the stripes to be the ground states, $\lambda$ must be also bounded from below. Indeed, when all 
$Z_i=-1(+1)$ are the same, then the $\lambda$-term contributes the energy of $2\lambda$ per site and the other 
term becomes 1D spin Heisenberg chains along the $a(b)$-axis. Their ground state has energy $-0.444$ per site. 
The total energy of the 1D Heisenberg phase is $-0.444+2\lambda$. It is higher than the stripe energy $-3/8$ 
when $\lambda>0.0345$. In conclusion, when $\lambda\in(0.0345,3/16)$ the SOE-RVB belongs to the ground state
manifold of $H_0$. 

Interestingly, the degeneracy of the stripes with different ${\cal N}$ can be removed by a simple perturbation
\be 
V=g\sum_i X_i.
\ee 
Here $X=|0\rangle\langle1|+|1\rangle\langle0|$ is an orbital Pauli matrix. A second order perturbative calculation 
shows that for $\lambda>0.0625$ each unpaired singlet adds a negative contribution to the energy of the ground state
making larger ${\cal N}$ more favorable.

\section{Kugel-Khomskii model}\label{sec:KKH}

A realistic version of the toy model (\ref{Htoy}) is a 2D Kugel-Khomskii model.
The perturbation theory for a Mott insulator with active $e_g$ orbitals 
in the regime of $t\ll U$ leads to the spin-orbital model \cite{Ole00}
\begin{eqnarray}
\label{eq:hamik}
&&
H  =  -\frac{1}{2}J\sum_{\langle ij\rangle||\gamma}\left\{ 
\left(r_{1}\,\Pi_{t}^{(ij)}+r_{2}\,\Pi_{s}^{(ij)}\right)
\left(\frac{1}{4}-\tau_{i}^{\gamma}\tau_{j}^{\gamma}\right)+\right.\nonumber \\
&&
\left.\left(r_{3}+r_{4}\right)\Pi_{s}^{(ij)}
\left(\frac{1}{2}-\tau_{i}^{\gamma}\right)
\left(\frac{1}{2}-\tau_{j}^{\gamma}\right)\right\} 
- E_{z} \sum_{i}\tau_{i}^{c},  
\end{eqnarray}
The model describes the spin-orbital superexchange in K$_2$CuF$_4$ 
\cite{Mos04} with $J=4t^2/U$.
The terms proportional to $r_1\equiv 1/(1-3\eta)$, $r_2=r_3\equiv 1/(1-\eta)$, and $r_4\equiv 1/(1+\eta)$
refer to the charge excitations 
to the upper Hubbard band \cite{Ole00} which 
depend on Hund's exchange parameter 
$
\eta=\frac{J_{H}}{U}.
$
The spin projection operators 
\begin{equation}
\Pi_{s}^{(ij)}=\left(\frac{1}{4}-{\bf S}_{i}\cdot{\bf S}_{j}\right),\hskip.5cm
\Pi_{t}^{(ij)}=\left(\frac{3}{4}+{\bf S}_{i}\cdot{\bf S}_{j}\right).
\label{eq:proje}
\end{equation}
select, respectively, a singlet or triplet configuration for spins $S=1/2$ on the bond 
$\langle ij\rangle$. 

Here $\tau_{i}^{\gamma}$ act in the subspace of $e_{g}$ orbitals occupied by a hole $\{|x\rangle,|z\rangle\}$, with 
$|z\rangle\equiv(3z^2-r^2)/\sqrt{6}$ and 
$|x\rangle\equiv(x^2-y^2)/\sqrt{2}$ --- they can be expressed in terms of 
Pauli matrices as
\cite{Ole00}: 
\begin{equation}
\tau_{i}^{a(b)}\equiv\frac{1}{4}
\left(-X_i \pm \sqrt{3} Z_i \right),
\quad\tau_{i}^{c}=\frac{1}{2}\,X_i.
\label{eq:odefs}
\end{equation}
Finally, $E_z$ is the crystal field splitting of two $e_g$ orbitals, induced by the lattice geometry or pressure.

At small enough $\eta$ and a suitable $E_z<0$, the PVB state was argued \cite{Brz12} to be a ground 
state of (\ref{eq:hamik}). At $\eta=0$ its energy is
\be 
\langle {\rm PVB} |H| {\rm PVB} \rangle=-\frac{3(23+4\sqrt3)}{128}JN=-0.7014JN,
\ee
where $N$ is the number of lattice sites. It is straightforward, though cumbersome, to check that any striped covering 
has the same energy as the PVB state.

\section{Conclusion}\label{sec:conclusion}

The spin-orbital entangled resonating-valence-bond state was shown to be a quantum superposition
of striped spin-singlet coverings that conserve a topological quantum number equal to the 
number of unpaired singlets in a periodic stripe. Its correlations are critical. The SOE-RVB state 
is a ground state of a simple spin-orbital Hamiltonian. It has the same energy as the PVB state in 
the realistic 2D Kugel-Khomskii Hamiltonian.

\acknowledgements
This work was supported by the Polish National Science Center (NCN) under Project DEC-2013/09/B/ST3/01603.

\section*{
Appendix: \\
correlations along a stripe for a low density of unpaired singlets
}

For a given stripe covering $C$, the correlator $\langle Z_l Z_{l+R} \rangle_C$ along a stripe depends on a number $n$
of unpaired singlets between the sites $l$ and $l+R$ as
\bea
\langle Z_l Z_{l+R} \rangle_C &=& 
\begin{cases}
1  & \text{if } R+n+e = 0,1,4,5,8,9,\dots\\
-1 & \text{if } R+n+e = 2,3,6,7,10,11,\dots
\end{cases}.
\nonumber
\eea
Here $e=0,1$ depending on the highest site number $l'\leq l$ occupied by an unpaired singlet. $e=1(0)$ when $l-l'$ is 
even(odd). Since in the dilute limit $e$ is equally likely to be $0$ or $1$, 
then only contributions of coverings with even $R+n$ survive in $\langle Z_l Z_{l+R} \rangle$. 
In other words, a surviving contribution of a covering $C$ is 
\bea
\langle Z_l Z_{l+R} \rangle_C^{\rm surv} &=& 
\begin{cases}
1  & \text{if } R+n = 0,4,8,12,\dots\\
-1 & \text{if } R+n = 2,6,10,14,\dots\\
0  & \text{otherwise}
\end{cases}\nonumber\\
&=& 
\cos\left[\frac{\pi}{2}(R+n)\right].
\eea
The correlator is an average over $n$:
\be 
C_R = \langle Z_l Z_{l+R} \rangle =  \sum_{n=0}^{R-1} p_n \cos\left[\frac{\pi}{2}(R + n)\right].
\label{corr}
\ee
Here $p_n$ is a probability distribution for $n$.

\begin{figure}[t]
\begin{center}\includegraphics[width=8cm]{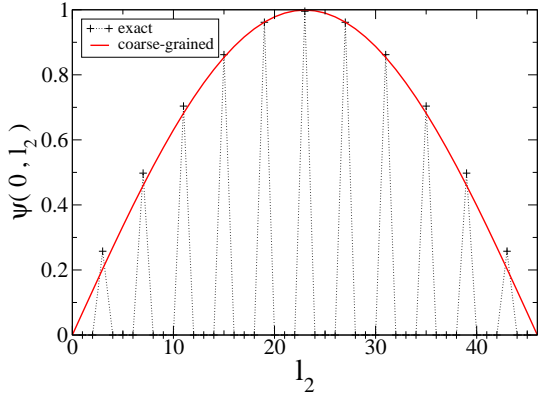}\end{center} 
\vspace{-0.7cm}
\caption{
The leading eigenvector of the striped transfer matrix $T_S$ for ${\cal N}=2$ unpaired singlets on $L=46$ sites. 
The plots compare the exact leading eigenfunction $\psi(0,l_2)$ of the exact $T_S$ in Eq. (\ref{TS}) with its 
coarse-grained counterpart in Eq. (\ref{TGS}). 
Both functions are in arbitrary units. 
The coarse-grained eigenfunction captures accurately the envelope of the exact one.
}
\label{sinpsi} 
\end{figure}

The striped transfer matrix (\ref{TS}) is written in the language of hard-core bosons. As long as only coarse-grained features of 
its leading eigenvector, like the distribution $p_n$ between distant sites, are concerned in the regime of low density, the constraint 
imposed by the projector $P$ can be relaxed. It affects only fine details of the wave function on the scale comparable to the lattice 
constant. By the same token, the next-NN hopping can be replaced by a NN hopping so that the transfer matrix becomes
\be 
T_{\rm S}~\approx~:\left( \sum_{l=1}^L c^\dag_{l+1} c_l + {\rm h.c.} \right)^{\cal N}:~.
\label{tildeTS}
\ee 
An application of this $T_{\rm S}$ to a Fock state with ${\cal N}$ particles creates a superposition of Fock states 
with each particle shifted by $\pm1$ lattice site with respect to its original location. In other words, a repeated application
of $T_{\rm S}$ generates a ${\cal N}$-particle random walk of particles constrained not to occupy the same site.
The ${\cal N}$-particle probability distribution describing the stochastic process converges with time to a stationary one:
\be 
\psi(l_1,...,l_{\cal N})~=~\prod_{i<j}\sin\left(\frac{\pi|l_i-l_j|}{L}\right).
\label{TGS}
\ee
An accuracy of this coarse-grained approximation can be appreciated in Figure \ref{sinpsi}. 
Interestingly,
this coarse-grained leading eigenvector of $T_S$ is the ground state of the 1D Tonks gas of hard-core bosons.

In the dilute regime, $p_n$ is a probability to find $n$ particles between the sites $l$ and $l+R$ in the state (\ref{TGS}).
$p_n$ does not change when we replace the hard-core bosons with fermions \cite{1Drev} and the state (\ref{TGS})
with a Fermi sea of ${\cal N}$ particles on $L$ sites. For fermions the characteristic function of $p_n$,
\be
\chi_{\lambda} = \sum_{n=0}^{R-1} p_n e^{i \lambda n},
\label{char_func1} 
\ee
is known \cite{Fisher-Hartwig} for a large block of length $R\to\infty$ in an infinite system:
\bea
\chi(\lambda) &=& G^2(1+\lambda/2 \pi) G^2(1-\lambda/2 \pi) \times \nonumber\\
              & & e^{i\lambda\rho R} \left(2 R \sin \pi\rho \right)^{-\lambda^2/2\pi^2} .
\label{char_func2}
\eea
Here $\rho$ is a particle density and $G$ denotes the G-Barnes function.

\begin{figure}[t]
\begin{center}\includegraphics[width=8cm]{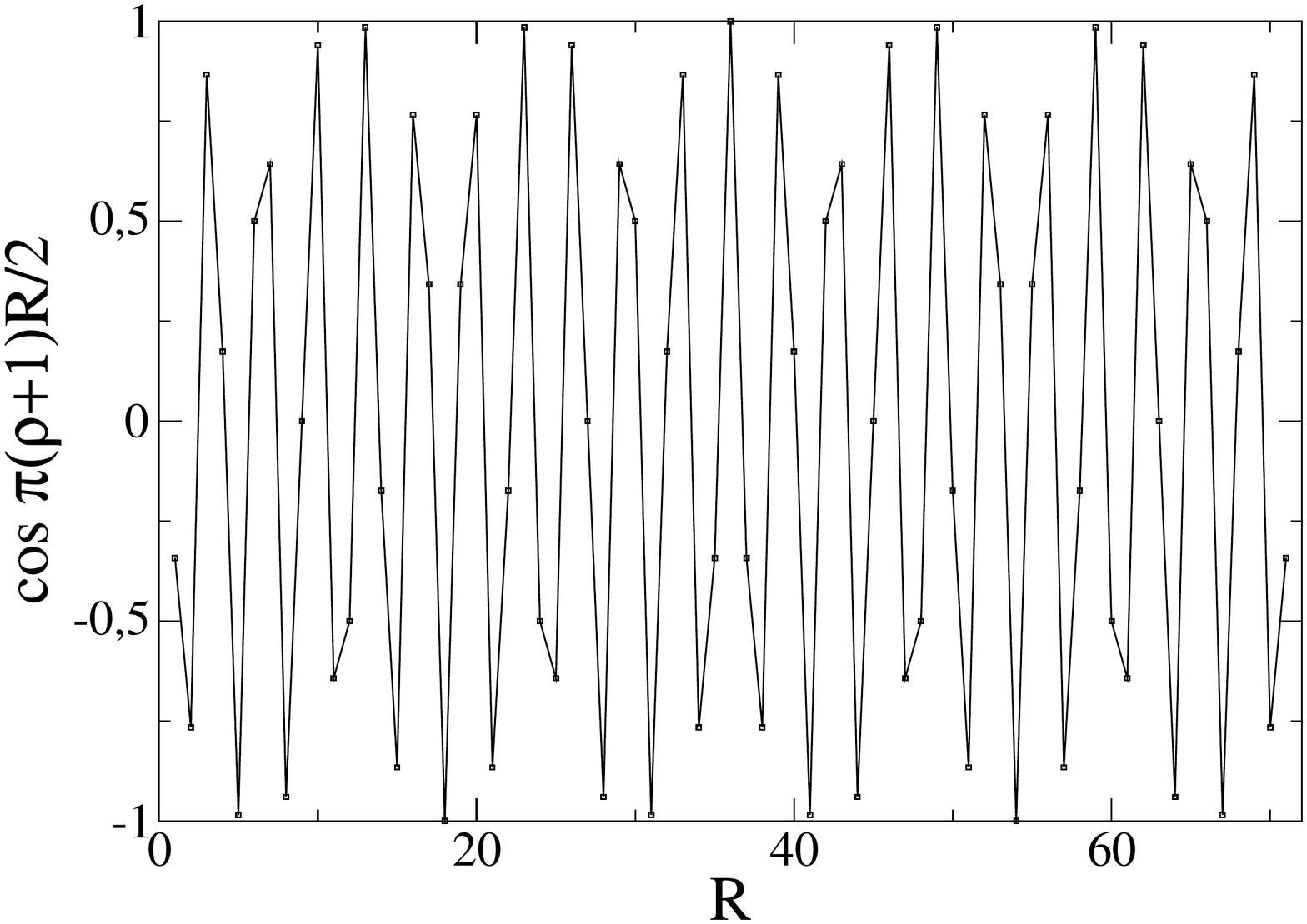}\end{center} 
\vspace{-0.7cm}
\caption{
The oscillating part of the analytic correlator (\ref{CRlow}) along a periodic stripe of length $L=72$
for ${\cal N}=16$ unpaired singlets and a singlet density $\rho={\cal N}/L=16/72$. These parameters
are the same as in Fig. \ref{XXvert}. In both Figures the fast oscillations are the same.
}
\label{XXvertanal} 
\end{figure}

In general $R=4r+\delta$, where $r$ is an integer and $\delta=0,1,2,3$ is a remainder. Since in the dilute limit
the probabilities $p_n$ for $n$ close to $R-1$ are exponentially small, we can proceed with Eq. (\ref{corr}) as
follows:
\bea 
C_R &=& \sum_{n=0}^{4r+\delta-1} p_n \cos\frac{\pi}{2}(4r+\delta+n) \nonumber\\
    &\approx & \sum_{n=0}^{4r-1} p_n \cos\frac{\pi}{2}(\delta+n)    \nonumber\\
    &=& \frac{\chi_{\frac{\pi}{2}}+\chi_{-\frac{\pi}{2}}}{2}  \cos\frac{\pi\delta}{2}-
        \frac{\chi_{\frac{\pi}{2}}-\chi_{-\frac{\pi}{2}}}{2i} \sin\frac{\pi\delta}{2} \nonumber\\
    &\sim & \left(\rho 4r\right)^{-1/8}
            \left(
            \cos\frac{\pi\rho 4r}{2}\cos\frac{\pi\delta}{2}-    
            \sin\frac{\pi\rho 4r}{2}\sin\frac{\pi\delta}{2}            
            \right)   \nonumber\\
    &=& \left(\rho 4r\right)^{-1/8}
        \cos\frac{\pi}{2}(\rho 4r+\delta) \nonumber\\
    &\approx &
        \left(\rho R\right)^{-1/8}
        \cos\frac{\pi}{2}(\rho+1)R.     
\label{CRlow}              
\eea 
The correlator along a stripe decays slowly with the exponent $1/8$. This may explain the modest decay with $R$ 
observed in Fig. \ref{XXvert}. 

Remarkably, the low density formula (\ref{CRlow}) captures correctly the fast oscillations in Fig. \ref{XXvert}. 
Indeed, in Fig. \ref{XXvertanal} the function $\cos\frac{\pi}{2}(\rho+1)R$ is shown for a stripe of length $L=72$ 
and the density of unpaired singlets $\rho=16/72$ just like in Fig. \ref{XXvert}. The fast oscillations in both 
Figures are the same.



\end{document}